\newcommand{\be}{\begin{equation}}
     \newcommand{\ee}{\end{equation}}
     \renewcommand{\ge}{\; \gtrsim \;}

\documentclass{article} 
\usepackage{amsfonts,amssymb}

\begin{document}

      \begin{center}
      {\bf Uncertainty in Measurements of Distance \\}
      \vspace{0.5cm}
      {\em John C.\ Baez\\}
      \vspace{0.3cm}
      {\small Department of Mathematics, University of California\\ 
      Riverside, California 92521 \\
      USA\\ }
      \vspace{0.5cm}
      {\em S.\ Jay Olson\\}
      \vspace{0.3cm}
      {\small Department of Physics, University of California\\ 
      Riverside, California 92521 \\
      USA\\ }
      \vspace{0.3cm}
      {\small email: baez@math.ucr.edu, olson@math.ucr.edu \\}
      \vspace{0.3cm}
      {\small January 9, 2002 \\ }
      \end{center}

\begin{abstract} 
\noindent 
Ng and van Dam have argued that quantum theory and general relativity
give a lower bound $\Delta \ell \ge \ell^{1/3} \ell_P^{2/3}$ on the
uncertainty of any distance, where $\ell$ is the distance to be measured
and $\ell_P$ is the Planck length.  Their idea is roughly that to
minimize the position uncertainty of a freely falling measuring device
one must increase its mass, but if its mass becomes too large it will
collapse to form a black hole.  Here we show that one can go below the
Ng--van Dam bound by attaching the measuring device to a massive elastic
rod.  Relativistic limitations on the rod's rigidity, together with the
constraint that its length exceeds its Schwarzschild radius, imply that
zero-point fluctuations of the rod give an uncertainty $\Delta \ell \ge
\ell_P$.

\end{abstract}

\section{Introduction}

It has long been believed that quantum gravity effects become important
at distances comparable to the Planck length, $\ell_P$, and a number
of arguments have been presented to support this idea \cite{AS,MTW}.
To show this sort of thing, one mainly needs to show that
gravity effects become important at {\it some} fixed length scale
depending only on the constants $c$, $G$ and $\hbar$.  
Dimensional analysis does the rest, since $\ell_P$
is the only quantity with dimensions of length that one can construct 
from these constants.  However, in situations where a
second length scale becomes relevant, one cannot use dimensional
analysis to settle all controversies.  For example, Ng and van Dam
\cite{NvD} have recently argued that quantum gravity effects cause
surprisingly large uncertainties in the measurement of a large distance
$\ell$, namely
\[
\Delta \ell \ge \ell^{1/3} \ell_P^{2/3} 
\]
where the symbol $\ge$ means that we are ignoring a constant
factor of order unity.  Amelino-Camelia \cite{Amelino} has 
gone even further, arguing that
\[
\Delta \ell \ge \ell^{1/2} \ell_P^{1/2} .
\]
Uncertainties on this scale are on the brink of being experimentally
detectable, lending extra interest to the issue.  However, 
in what follows, we reanalyze the Ng--van Dam thought experiment and show
that by modifying its design we can dramatically reduce the uncertainty
of distance measurements.  Our modified thought experiment gives
\[
\Delta \ell \ge \ell_P .
\]

\section{Ng--van Dam Thought Experiment}

The elements of the Ng--van Dam thought experiment are straightforward,
and the aim is to show that through a simple application of the
uncertainty principle, together with limits imposed by general
relativity, we are led to the conclusion that a fundamental distance
uncertainty arises that may be far larger than the Planck scale.

The argument proceeds as follows.  First consider two nearby objects in
free fall approximately at rest relative to one another: an observer
consisting of a clock and light emitter, and a mirror.  If the observer
wants to know the distance to the mirror, he may simply emit a burst of
light, wait a time $t$ for the light to return, and conclude that the
mirror is a distance $\ell = ct/2$ away.

Now we are interested in the uncertainty of this measurement.  
Following an argument due to Wigner \cite{SW,Wigner} we treat the clock as
a free quantum mechanical particle and impose the uncertainty condition 
$\Delta q \Delta p \ge \hbar$.  Writing $\Delta p = m \Delta v$ where $m$
is the mass of the clock, we thus obtain the following bound on the
uncertainty of the clock's position at time $t$:
\be
\begin{array}{ccl}
\Delta q(t) &=& \Delta (q + tv)  \\ 
            &=& \sqrt{(\Delta q)^2 + (t \Delta v)^2} \\ 
            &\ge& \Delta q + \frac{\hbar t}{m \Delta q} .
\end{array}
\label{1}
\ee
To minimize the position uncertainty at time $t$, we find that the
optimal position uncertainty at time zero should be $\Delta q =
\sqrt{\hbar t / m}$.  Plugging this back into equation (\ref{1}), we
find the minimum uncertainty at time $t$ to be 
\be
\Delta q(t) \ge \sqrt{\hbar t / m} .
\label{2}
\ee
It is also convenient to write this in terms of the distance
to be measured and the Compton wavelength of the clock, 
$\ell_C = \hbar / m c$:
\be
\Delta q(t) \ge \ell^{1/2} \ell_C^{1/2} .
\label{3}
\ee
This uncertainty in the position of the clock contributes to the
uncertainty in $\ell$, the distance between the clock and mirror.  We can
ignore the uncertainty in the position of the mirror, which behaves
similarly, and obtain this lower bound on $\Delta \ell$:
\be
\Delta \ell \ge \ell^{1/2} \ell_C^{1/2} .
\label{4}
\ee

So far we have only considered the effects of quantum mechanics and the
speed of light, with no mention of the effects of general relativity. 
Next, Ng and van Dam consider the details of the clock itself.  They
take the clock to consist of two parallel mirrors a distance $d$ apart,
and consider a tick of the clock to be the time $2 d / c$ that it takes
light to travel back and forth between them.  Since we now have the
length scale $d$ and the mass scale $m$ of the clock, we can now begin
to consider general relativity effects.  In particular, Ng and van Dam
assert that the size of the clock, $d$, must be larger than its
Schwarzschild radius $\ell_S = Gm/c^{2}$.  If the tick of the clock is a
lower bound on the accuracy of its time measurements, this requirement
implies that 
\be
\Delta \ell \ge \ell_S.
\label{5}
\ee

Finally, squaring the uncertainty from equation (\ref{4}) and
multiplying the result by equation (\ref{5}), we obtain $(\Delta \ell)^3
\ge \ell \ell_C \ell_S$.  Note that $\ell_C \ell_S$ is equal to
$\ell_P^2$, the Planck length squared.  Thus the primary result obtained
from the Ng--van Dam thought experiment is that the minimum uncertainty
in this kind of measurement satisfies
\be 
\Delta \ell \ge \ell^{1/3} \ell_P^{2/3},
\label{6} 
\ee 
a bound depending only on the distance $\ell$ to be measured
and the Planck length.

Ng and van Dam have suggested that this uncertainty is inherent in any
distance measurement, and that it may be apparent in the latest
generation of interferometers designed for detecting gravitational
waves.  As explained in Section 3, we believe that this
conclusion is unwarranted, and that changes in the details of the clock
used in the thought experiment will have an effect on the final
uncertainty result, and thus lead to a breakdown of the Ng--van Dam
uncertainty as a fundamental property of distance measurement.
 
Before turning to this we should comment a bit on the Wigner clock
thought experiment.   Our derivation of equation (\ref{2}) is rigorous
{\it if} in the second step we assume that the fluctuations in $q$ and
$p$ are uncorrelated, or more precisely, that the expectation value
$\langle pq + qp \rangle$ equals $2\langle p \rangle \langle q \rangle$.
This is true if, for example, the wavefunction for the clock's center of
mass is initially Gaussian.   If we allowed the wavefunction to be
arbitrary, we could arrange for $\Delta q(t)$ to be arbitarily small
at some chosen time $t$, simply by taking a wavefunction close to a delta
function at time $t$ and evolving it backwards to get the wavefunction
at time zero.  However, for our purposes this counts as `cheating', since
we can only reduce $\Delta q(t)$ arbitrarily this way if we already know
the time $t$ that is to be measured.  

\section{A Modified Thought Experiment}

As was pointed out by Adler, Nemenman, Ovenduin, and Santiago
\cite{ANOS}, the result of the above thought experiment is only valid
when the clock evolves according to the free Schr\"odinger equation, and
this need not be the case.  Ng and van Dam \cite{NvD2} have, in turn,
asserted that if the clock is bound through some potential, then it must
be bound to something, and that this something can then be considered as
a part of the clock.  We will explore this line of reasoning through a
specific example, and argue that attaching the clock to an external
object does indeed affect the final uncertainty result.

For simplicity, let us take the clock to be attached to one end of a rod
in free fall.  We assume this rod has equilibrium length $x$, elastic
constant $Y$, and mass $m$.  We take the mass of the clock to be
negligible compared to $m$.  It is thus the rod's mass, rather than that
of the clock, which limits the spreading of the clock's wavefunction
with the passage of time.  By making $m$ large we can make this spread
as slow as we like.  However, special relativity puts a bound on the
elastic constant $Y$, since a very rigid medium would allow sound waves
to propagate with speed greater than that of light.  This means that the
ends of the rod will undergo zero-point fluctuations which also
contribute to the uncertainty of any position we measure using the
clock.  

To calculate these, recall that the speed of sound along the rod
will be $\sqrt{Yx/m}$.  Since this must be less than $c$, we have
\be
Y \le mc^2/x .
\label{7}
\ee
Now, a one-dimensional rod with one end attached to an immovable wall
can perform oscillations about its equilibrium length $x$ exactly as a
harmonic oscillator with spring constant $k = Y/x$ and oscillator mass
$m_{osc} = m/3$.  Thus, a rod in space will have a mode of oscillation
that allows each end to perform harmonic oscillations about its
respective equilibrium position with equivalent spring constant $k =
2Y/x$ and oscillator mass $m_{osc} = m/6$.  Treating the problem
quantum-mechanically, the ground state of such an oscillator has
position uncertainty 
\be
\Delta x = \sqrt{\frac{\hbar}{2(km_{osc})^{1/2}}} 
         = \sqrt{\frac{\hbar}{2} \left(\frac{3x}{mY}\right)^{1/2}}
\label{8}
\ee
Using equation (\ref{7}) and dropping factors of order unity, we 
obtain
\be
\Delta x \simeq \sqrt{\hbar x /mc} 
\label{9}
\ee
There are other states where $\Delta x$ is smaller at one moment, but
not for the entire duration of the experiment, at least if the rod's
period of oscillation is short compared to the time $t$ which the clock
is to measure --- and if it were not, the rod would fail to slow the
spread of the clock's wavefunction.  We may thus take this value of
$\Delta x$ as an approximate lower bound on the uncertainty of the
clock's position with respect to rod's center of mass.  In terms of the
Compton wavelength of the rod, it follows that
\be
\Delta x \ge x^{1/2} \ell_C^{1/2}.
\label{10}
\ee

The rod's center of mass will also have an uncertainty in its position,
given by equation (\ref{3}).   Both this and $\Delta x$ will contribute
to the uncertainty of any distance we measure by sending a burst of
light from the clock and measuring the time it takes for the light to
return.  Assuming these uncertainties are uncorrelated, we obtain 
\be 
\Delta \ell \ge (\ell^{1/2} + x^{1/2})\, \ell_C^{1/2} . 
\label{11} 
\ee

Next we must rethink the requirements imposed by general relativity on
the clock--rod system.  The requirement made in the original Ng--van Dam
thought experiment was that the clock must be larger than its own
Schwarzschild radius.  While this is still true, it is not important
now, since we no longer need the clock to be heavy to prevent its wave
function from spreading with time.  Instead, the important relativity
consideration is that the rod be longer than its Schwarzschild radius:
\be
x \ge \ell_S
\label{12}
\ee
Although we do not have spherical symmetry in this example, the
Schwarzschild radius remains useful as a quick and dirty estimate of the
length scale at which the rod would collapse to form a black hole.
Indeed, the `hoop conjecture' \cite{Thorne} says roughly that {\it any}
system compressed within its Schwarzschild radius must give rise to a
singularity.

Plugging this lower bound on $x$ into equation (\ref{10}), we obtain
\begin{equation}
\Delta \ell \ge (\ell^{1/2} + \ell_S^{1/2})\, \ell_C^{1/2} .
\label{13}
\end{equation}
We can minimize this uncertainty by making the rod very heavy, so that
$\ell_C \to 0$, leaving us with
\begin{equation}
\Delta \ell \ge \ell_P.
\label{14}
\end{equation}

It is interesting to compare this argument to that given by Ng and van
Dam.  First, unlike their argument, ours does not assume any limitation
on the clock's accuracy solely due to its size.  Second, while their
result was obtained by multiplying two independent lower bounds on
$\Delta \ell$, one from quantum mechanics and the other from general
relativity, ours arises from an interplay between competing effects.  On
the one hand, we wish to make the rod as heavy as possible to minimize
the quantum-mechanical spreading of its center of mass.  To prevent it
from becoming a black hole, we must also make it very long.  On the
other hand, as it becomes longer, the zero-point fluctuations of its
ends increase, due to the relativistic limitations on its rigidity.  We
achieve the best result by making the rod just a bit longer than its own
Schwarzschild radius.

We should also mention and rebut a possible objection to our argument.
One could try to eliminate the effect of zero-point fluctuations on the
position of the clock by attaching the clock to the midpoint of the rod
rather than the end.  This would indeed eliminate the effect of the
vibrational mode we have been considering.  However, there are other
modes in which the midpoint of the rod oscillates.  If we take these
into account, some of the numbers in our formulas change, but we again
find that $\Delta \ell$ is greater than $\ell_P$ times some constant of
order unity.

\section{Conclusions}

It appears that by attaching the clock used in the Ng--van Dam thought
experiment to a heavy object, the surprisingly large uncertainties
arising in distance measurements can be avoided.  It is thus our opinion
that these large uncertainties cannot be a fundamental feature of the
union of quantum mechanics and general relativity.  

Following Ng and van Dam, most recent papers on this topic suggest laser
interferometry as a possible way to actually detect the uncertainties in
measurements of distance caused by quantum gravity effects
\cite{Amelino,ANOS,AS,NvD,NvD2}.   While this is tempting, it is worth
noting that interferometers do not measure distances so much as periodic
changes in distance.  Indeed, the design of an interferometer is very
different from the Ng--van Dam thought experiment or our modified
version, so any attempt to detect quantum gravity effects at LIGO will
require a separate analysis.

\end{document}